\documentclass[seceq,supplement]{ptptex}
\usepackage{wrapft}
\usepackage{graphicx}

\pubinfo{No.~158, 2005}
\notypesetlogo                       

\title{Observed Properties of Exoplanets: \\
Masses, Orbits, and Metallicities
}

\author{
Geoffrey \textsc{Marcy},$^{1,}$\footnote{E-mail: gmarcy@berkeley.edu} R. Paul \textsc{Butler},$^{2}$ Debra
\textsc{Fischer},$^{3}$\\ Steven \textsc{Vogt},$^{4}$ Jason
T. \textsc{Wright},$^{1}$ Chris G. \textsc{Tinney}$^{5}$ and Hugh R. A. \textsc{Jones}$^{6}$}

\inst{
$^1$Astronomy Department, University of California,
Berkeley, CA 94720, USA \\
$^2$Department of Terrestrial Magnetism, Carnegie Institution of
Washington,\\ 5241 Broad Branch Rd NW,  Washington DC  20015-1305, USA\\
$^3$Department of Physics and Astronomy, San Francisco State
University, \\
San Francisco, CA 94132, USA \\
$^4$UCO/Lick Observatory, 
University of California at Santa Cruz, \\ Santa Cruz, CA 95064, USA \\
$^5$Anglo-Australian Observatory, PO Box 296, Epping 1710, Australia \\
$^6$Centre for Astrophysics Research, University of Hertfordshire,\\
Hatfield, AL10 9AB, UK
}

\abst{We review the observed properties of exoplanets found by the
Doppler technique that has revealed 152 planets to date.  We focus on
the ongoing 18-year survey of 1330 FGKM type stars at Lick, Keck, and
the Anglo-Australian Telescopes that offers both uniform Doppler
precision (3 {\mbox{m s$^{-1}$}}) and long duration.  The 104 planets
detected in this survey have minimum masses (M$\sin i$) as low as 6
$M_{Earth}$, orbiting between 0.02 and 6 AU.  The core-accretion model
of planet formation is supported by four observations: \ 1) The mass
distribution rises toward the lowest detectable masses, d$N$/d$M
\propto M^{\rm -1.0}$. \ 2) Stellar metallicity correlates strongly
with the presence of planets. \ 3) One planet (1.3 M$_{\rm Sat}$) has
a massive rocky core, M$_{\rm Core}\approx $70 M$_{\rm Earth}$.  4) A
super-Earth of $\sim$7 M$_{\rm Earth}$ has been discovered.  The
distribution of semi-major axes rises from 0.3 -- 3.0 AU (d$N$/d$\log
a$) and extrapolation suggests that $\sim$12\% of the FGK stars harbor
gas-giant exoplanets within 20 AU.  The median orbital eccentricity is
$<e>$=0.25, and even planets beyond 3 AU reside in eccentric orbits, 
suggesting that the circular orbits in our Solar System
are unusual.  The occurrence ``hot Jupiters'' within 0.1 AU of FGK
stars is 1.2$\pm$0.2\%.  Among stars with one planet, 14\% have at
least one additional planet, occasionally locked in resonances.
Kepler and COROT will measure the occurrence of earth-sized planets.
The Space Interferometry Mission (SIM) will detect planets with masses
as low as 3 $M_{\rm Earth}$ orbiting within 2 AU of stars within 10
pc, and it will measure masses, orbits, and multiplicity.  The
candidate rocky planets will be amenable to follow-up spectroscopy by
the ``Terrestrial Planet Finder'' and Darwin.}

\begin{document}

\maketitle

\section{Introduction}

In the past 10 years, 152 exoplanets have been discovered orbiting 131
normal stars by using the Doppler technique to monitor the
gravitational wobble induced by a planet, as previously summarized
\cite{Marcy_Paris_04, Mayor04}.  Multiple planet systems have been
detected around 17 of the 131 planet-bearing stars, found by
superimposed multiple Doppler periodicities \cite{Vogt05,Mayor04}.
  Remarkable statistical properties have
emerged from the 152 planets:
\medskip
\begin{itemize}
\item{Planet mass distribution: d$N$/d$M \propto M^{-1.0}$ (Fig. 1)}
\item{$>$7\% of stars have giant planets within 5 AU, most beyond 1 AU (Fig. 2)}
\item{Hot Jupiters ($a < 0.1$ AU) exist around 1.2\% of FGK stars}
\item{Eccentric Orbits are common, with a median of $\langle e\rangle$ = 0.25 (Fig. 3)}
\item{Planet occurrence rises rapidly with stellar metallicity (Fig. 6)}
\item{Multiple planets are common, often in resonant orbits (Fig. 7)}
\end{itemize}
\medskip

Four planets of extraordinarily low mass have been found.  Three have
Neptune-like masses of $M{\sin i}$ of 21, 15, and 18 $M_{\rm Earth}$
orbiting host stars, GJ~436, 55~Cancri, and HD~190360, respectively
\cite{Butler04,McArthur04,Vogt05}.  The fourth planet is likely the
first ``super-Earth'' with $M{\sin i}$ = 6.0 M$_{\rm Earth}$ and
$P$=1.94 d, orbiting the star, GJ 876.  Apparently, planet formation
can populate the mass range between that of Uranus and Earth.

The first direct image of an exoplanet has finally occurred with the
VLT/NACO and HST/NICMOS images of 2M1207 and its companion separated
by 773 mas (54 AU projected separation) \cite{Chauvin04,Schneider04}.
At the likely age (8 Myr) of this system in the TW Hydrae association,
the IR photometry implies a mass of 2--5 $M_{\rm Jup}$ based on
atmospheric models of such young, warm planets
\cite{Burrows97,Baraffe02}.  The second epoch HST observations
(Schneider, private communication) confirm that the companion is bound
to the primary, rendering it the first planetary mass companion ever
imaged.

To date, 7 planets are known that cross the disk of their star, 5 found
photometrically by the dimming of the star \cite{Alonso04, Torres05,
Bouchy05}.  However, the two closest stars with transiting planets (HD 209458, HD
149026) were found first by the Doppler method \cite{Henry00,
Charbonneau00, Sato 05}.  The fractional dimming of the star's flux
gives a direct measure of the radius of the planet relative to the
stellar radius that is determined from stellar modelling.  The edge-on
orbit and Doppler measurements give the planet mass.  The resulting
densities of these planets are in the range of 0.2--1.4 g/cm$^3$
verifying the expectation that the planets are gaseous (albeit with
{\it liquid} metallic hydrogen interiors \cite{Henry00,
Charbonneau00}.  During transit, starlight passing through the
planet's atmosphere has allowed detection of its constituents, notably
sodium and hydrogen \cite{Charbonneau01, Vidal-Madjar04}.  During
eclipse of the planet by the star, measurements of the diminished
infrared flux in narrow bands permits assessment of the atmospheric
temperature and other atmospheric constituents such as water vapor and
methane \cite{Deming05, Charbonneau05}.  From precise Doppler
measurements, including the Rossiter effect (in which the planet
blocks a portion of the rotating star's hemisphere, causing a net
Doppler shift), the tidal heating has been shown to be negligible,
leaving the large radius of HD~209458 unexplained \cite{Laughlin05,
Winn05}.  Most remarkable is HD~149026 that has only 1.21 M$_{\rm
Sat}$ but a high density of 1.4 g/cm$^3$ (twice Saturn's),
implying the existence of a massive rocky core of 70 M$_{\rm
Earth}$ \cite{Sato05}.

The standard theory for the formation of gas giant planets is the
``core-accretion'' model that begins with dust particles colliding and
growing within a protoplanetary disk to form rock-ice planetary cores
\cite{Greenberg78, Wetherill89, Aarseth93, Lissauer95, Levison98,
Kokubo98, Kokubo00, Kokubo02, Hayashi81, Ikoma00}.  If the core
becomes massive enough while gas remains in the disk, it
gravitationally accretes nearby gas, acquiring an extended gaseous
envelope \cite{Mizuno80, Bodenheimer86, Pollack96, Bodenheimer03}.
Support for the core-accretion model has come from HD~149026b 
with its massive rocky core
and implied high abundance of heavy elements, suggesting 
that core formation dominated any acquisition
of gas, as discussed by Sato et al. (2005).

Gas giants accrete most of the gas within their tidal reach
filling the Hill sphere around them with the heated,
gaseous envelope.  Further gas accretion is slowed both by the
diminishing amounts of remaining local gas and by the extended
envelope, leading to predicted growth times of 5--10 Myr.  This growth
time scale is uncomfortably longer than the observed $\sim$3 Myr
lifetime of the disks themselves \cite{Hartmann98, Haisch01}.
Therefore the original core accretion model suffers from a
planet-growth rate that is too slow.

This inadequacy in the core-accretion model has been addressed in two
ways.  Inward migration of type I \cite{Ward97, Lin00, Tanaka02,
Tanaka04} will bring giant planets to fresh, gas-rich regions of the
disk \cite{Alibert05, Armitage03}.  Moreover, improved opacities are
lower, allowing more rapid escape of radiation that speeds the shrinkage of
the envelope to accelerate the accretion of more gas.  The
resulting planet growth time scale is shortened to $\sim$1 Myr, well
within the lifetime of protoplanetary disks \cite{Alibert05,
Hubickyj04, Armitage03, Ida04a}.

Giant planets probably form preferentially {\em beyond 3 AU} where
their tidal reach permits accretion of large amounts of cool disk gas.
In contrast, we find that 1.2\% of stars harbor giant planets {\em
within 0.1 AU}, suggesting that the planets migrated inward. Strong
evidence of migration comes from the numerous resonances among the
multi-planet systems that suggests migrational settling into the
resonance traps \cite{Lee02, Chiang02, Vogt05, Rivera01, Laughlin01,
Kley05}. Migration may occur by two primary processes.  Planets may
lose energy and angular momentum to the disk (type I)
causing inward migration, or the disk gas may viscously accrete onto
the star, dragging planets with them (type II) \cite{Ward97,
Lin00, Bryden01, Trilling02, Chiang02, Dangelo03, Ida04b}

The origin of the orbital eccentricities remains poorly understood, as
interactions between planet and gaseous disk are thought to damp
eccentricities \cite{Ward93, Artymowicz93, Tanaka04}.  If so, {\it the
orbital eccentricities must arise after the major stage of gas
accretion.}  Subsequent gravitational interactions among planets and
between planets and the disk may cause the observed orbital
eccentricities, perhaps related to resonant interactions between
planets \cite{Bryden01, Rivera01, Laughlin01, Lee02, Chiang02, Ford03,
Ida04a, Ford05, Thommes03, Nelson02, Marzari02, Gozdziewski03}.

The properties of the masses and orbits of observed giant planets are
becoming well represented by the current models \cite{Alibert05,
Kley05} including the correlation of planets with metallicity
\cite{Ida04b} and with stellar mass \cite{Ida05} (see Ida \& Lin in
this volume).

\section{The Lick, Keck, and the Anglo-Australian planet search}

The determination of the statistical properties of giant planets
depends on a survey of planets that has well-understood detection
thresholds in both mass and orbital period.  We have carried out
precise radial velocity measurements of 1330 FGKM dwarfs at the Lick,
Keck~1, and Anglo-Australian telescopes.  The majority of stars had
their first high-quality measurement between 1995 and 1998, giving a
time coverage of $\sim$7--10 years thus far.  The target stars and
their properties are available \cite{Wright04a, Valenti05}, and were
drawn from the Hipparcos catalog \cite{ESA97} with criteria that they
have $B-V >$ 0.55, reside no more than 3 mag above the main sequence
(to avoid photospheric jitter seen in giants), and have no stellar
companion within 2 arcsec (to avoid confusion at the entrance slit).

The target list also includes 120 M dwarfs, located mostly within 10
pc with declination north of $-$30 deg \cite{Wright04a}.  For the
late-type K and M dwarfs, we restricted our selection to stars
brighter than $V=11$.  All slowly rotating stars are surveyed with a
Doppler precision of 3 {\mbox{m s$^{-1}~$}} to provide a uniform
sensitivity to planets.  Thus far, our Lick, Keck, and
Anglo-Australian surveys have revealed 104 planets orbiting 88 stars,
including 12 multi-planet systems.  The orbital elements and masses of
these exoplanets are regularly updated at:
\verb1http://exoplanets.org1 .

\section{Observed properties of exoplanets}

We derive the statistical properties of planets from the 1330 FGKM
target stars for which we have uniform precision of 3 {\mbox{m
s$^{-1}~$}} and at least 6 years duration of observations.  Detected
exoplanets have minimum masses, $M{\sin i}$, between 6 $M_{\rm
Earth}~$ and $\sim$15 {$M_{\rm Jup}$}, with an upper mass limit
corresponding to the (vanishing) tail of the mass distribution.  The
planet mass distribution is shown in Fig.~1 and follows a power law,
$dN/dM \propto M^{-1.05 ~}$ \cite{Marcy_Butler00,Marcy_DC_03} affected
very little by the unknown $\sin i$ \cite{Jorissen01}.  The paucity of
companions with $M{\sin i}$ greater than 12 $M_{\rm Jup}$ confirms the
presence of a ``brown dwarf desert'' \cite{Marcy_Butler00} for
companions with orbital periods up to a decade.
  
\begin{figure}
\includegraphics[width=14cm]{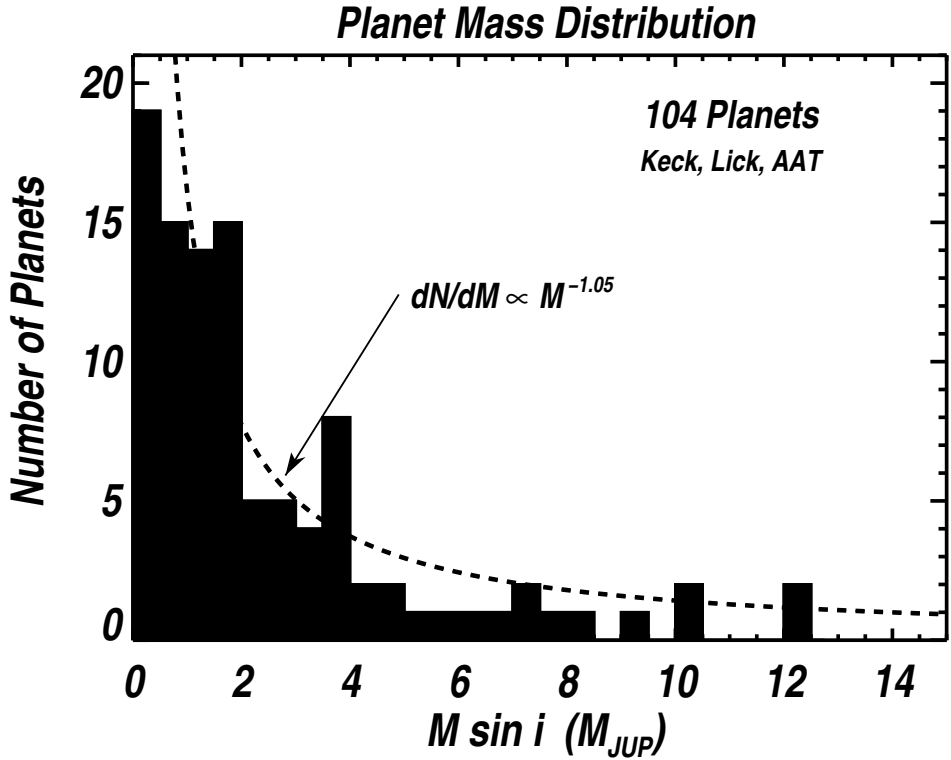}
\caption{The histogram of 104 planet masses ($M{\sin i}$) found in the
uniform 3 m\ s$^{-1}$ Doppler survey of 1330 stars at Lick, Keck, and
the AAT telescopes.  The bin size is 0.5 $M_{\rm Jup}$.  The
distribution of planet masses rises as $M^{-1.05}$ from 10 $M_{\rm
Jup}$ down to Saturn masses, with incompleteness at lower masses.}
\end{figure}

The 88 stars with planets among the 1330 target stars imply that the
fraction of stars harboring giant planets with $M < 13 M_{\rm Jup}$ within
5 AU is at least 88/1330 = 6.6\%.  This is no doubt a lower limit as planets
between 3--5 AU are not efficiently detected due to the limited
duration, 6--8 years, of our Doppler survey.  The 12 stars with
two or more planets implies an occurrence rate of at least 1\% for
systems with multiple giant planets.

The observed semimajor axes span the range 0.02--6.0 AU.  We have
found 16 planets that orbit within 0.1 AU, implying that such ``hot''
Jupiters exist around 16/1330 = 1.2$\pm$0.3\% of FGK main sequence
stars.\cite{Marcy_Paris_04, Jones04}   The number of planets increases
with distance from the star from 0.3 to 3 AU, as shown in Fig.~2 (in
logarithmic bins). A modest (flat) extrapolation suggests that a
comparable population of yet-undetected jupiters exists between 5--20
AU, bringing the occurrence of giant planets to roughly 12\% within 20
AU (Fig.~2).  Indeed, some 5\% of our stars show a long term trend in
velocity, suggestive of a planetary companion between 5 and 20 AU.

\begin{figure}
\includegraphics[width=14cm]{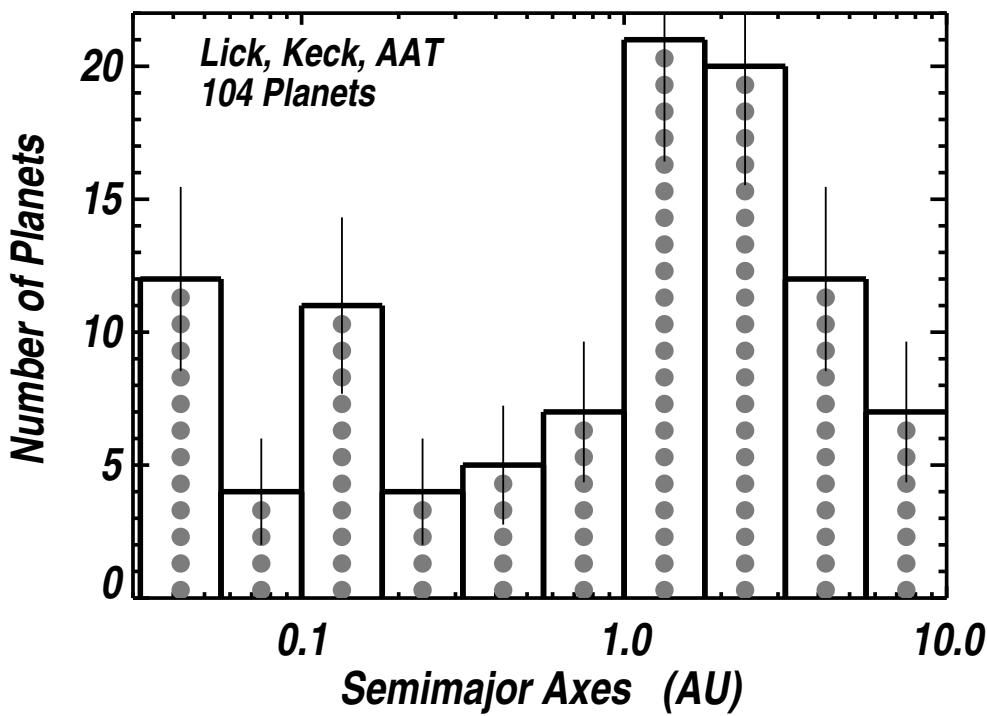}
\caption{Histogram of semimajor axes, $a$, of the 104 exoplanets
found from the Doppler survey at Lick, Keck, and the AAT Telescopes.
Note the equal logarithmic bins,$\Delta \log{a}$.  There are increasing
numbers of planets toward larger orbits beyond 0.5 AU.  The occurrence
of planets within 3 AU is 6.6\%.   There is increasing incompleteness
beyond 3 AU. Flat extrapolation from 3--20 AU
suggests that $\sim$12\% of all nearby FGK stars have a giant planet
with mass greater than Saturn.}
\end{figure}

The orbital eccentricities for the 104 detected exoplanets are plotted
versus semimajor axis in Fig.~3.  Apparently eccentricities span the
full available range, 0.0--1.0, but avoiding those with such small
periastron distances, $r_{\rm min}=a(1-e)<$0.1, that tidal
circularization would occur.  Indeed, planets orbiting within 0.1 AU
are all in nearly circular orbits, presumably due to tidal
circularization.

However, planets orbiting beyond 0.1 AU (i.e. not circularized) have a
median eccentricity of $\langle e\rangle=0.25$ with a standard
deviation of 0.19 .  Thus, the orbital eccentricities of giant planets
within 5 AU are considerably higher than those in our Solar System.
Remarkably, among the planets farthest from the host star, $a$ = 2--4
AU, there is no tendency for them to have small eccentricities.  The
indications from our velocity data suggest that exoplanets having $a
\approx$ 5.2 AU will also have non-zero eccentricities. It may be
years before we are able to distinguish definitively true
eccentricities at 5 AU from multiple planets.  In the upcoming years
we expect to discover a population of exoplanets at $a\approx$ 5.2 AU
that will allow direct comparison of cosmic eccentricities with that
of Jupiter, $e=0.048$.

\begin{figure}
\includegraphics[width=14cm]{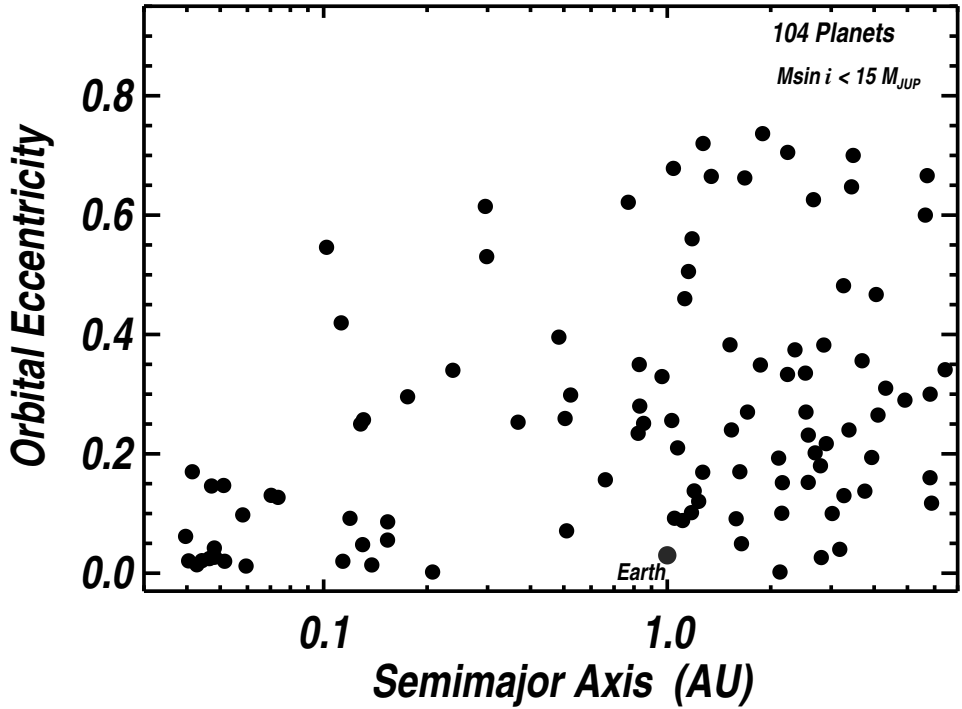}
\caption{Eccentricity vs semimajor axis, for the 104 planets
discovered in the Lick, Keck, and AAT Doppler survey.  Eccentricity
ranges from 0 to 0.8 and no decline in eccentricity is
observed beyond 3 AU.  It seems quite likely that giant planets at 5.2 AU 
also reside in eccentric orbits, in contrast to Jupiter in our
Solar System. }
\end{figure}

Our 104 planets can be examined for any relationship between planet
mass and semimajor axis.  We plot planet mass vs semimajor axis for
our 104 planets in Fig.~4.  The upper left region of the plot is
devoid of planets, indicating that massive planets rarely are found
close to the host star.  Our Doppler survey, with its uniform Doppler
precision of 3 m~s$^{-1}$ and duration of $\sim$8 years would have
easily found massive planets orbiting within 1 AU of target stars.
Thus, this paucity of massive, close-in planets is not a selection
effect and is statistically real \cite{Zucker03}.

However, we might ask if the mass distribution for planets orbiting
beyond 1 AU is actually different from those within 1 AU.
Interestingly, the planets both beyond and within 1 AU have a mass distribution
that increases toward lower masses (Fig.~4).   The small number of high
mass planets within 1 AU may be simply due to the small total number of
planets orbiting close in.  With our precision of 3 m~s$^{-1}$, planets beyond 1 AU
having low masses are certainly missed, both because they induce a
small wobble and because few orbits have transpired during our
program.  Thus while the small number of massive planets orbiting close to
host stars is real, there is no strong evidence that the mass
distribution is a function of orbital distance.

\begin{figure}
\includegraphics[width=14cm]{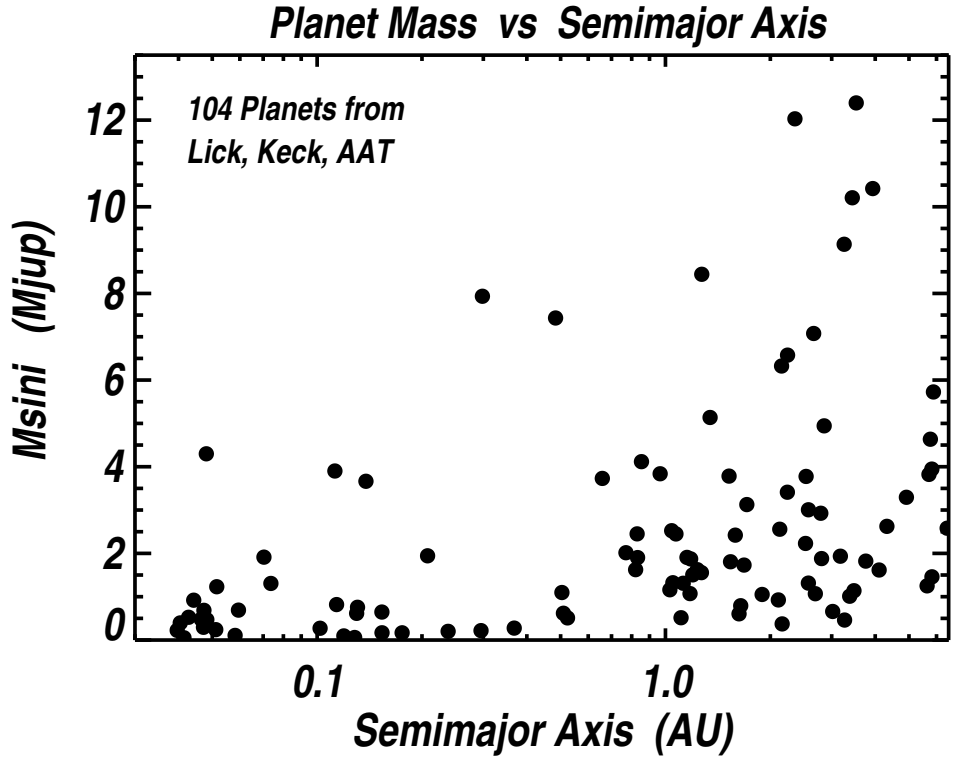}
\caption{Planet mass vs semimajor axis for the 104 planets
found at Lick, Keck, and the AAT.   There is a dearth of close-in planets
having high mass that cannot be a selection effect as our
survey would surely discover the massive, close-in planets.
However, the distribution of masses for close-in planets may be
similar to that for planets {\it beyond} 1 AU, considering the
difficulty in detecting low mass planets beyond 1 AU.}
\end{figure}

We may also consider the dependence of orbital eccentricity on planet
mass.  In Fig.~5, we plot eccentricity vs $M{\sin i}$ for the 104 planets
found in our Doppler survey, but we include only those 86 planets
orbiting with $a>$0.1 AU, to avoid those that may have been tidally
circularized.  Figure 5 shows no strong correlation between
eccentricity and planet mass.  However, the most massive planets,
notably those with $M{\sin i} >  5 \,M_{\rm Jup}$, exhibit systematically
higher eccentricities than do the planets of lower mass.  This cannot
be a selection effect nor can it be caused by errors because the most
massive planets (right half) induce the largest Keplerian amplitudes,
$K$, allowing accurate determination of eccentricity.

If planets form initially in circular orbits, the high eccentricities
of the most massive planets in Fig.~5 poses a puzzle.  Such massive
planets have the greatest inertial resistance to perturbations that
are necessary to drive them out of their initial circular orbits.  Yet
the massive planets reside mostly in orbits more eccentric than the
lower mass planets.  We remain puzzled that the most massive planets
have the highest orbital eccentricities.    {\em Perhaps massive planets 
formed by a process in which the orbits are not initially circular}.

\begin{figure}
\includegraphics[width=14cm]{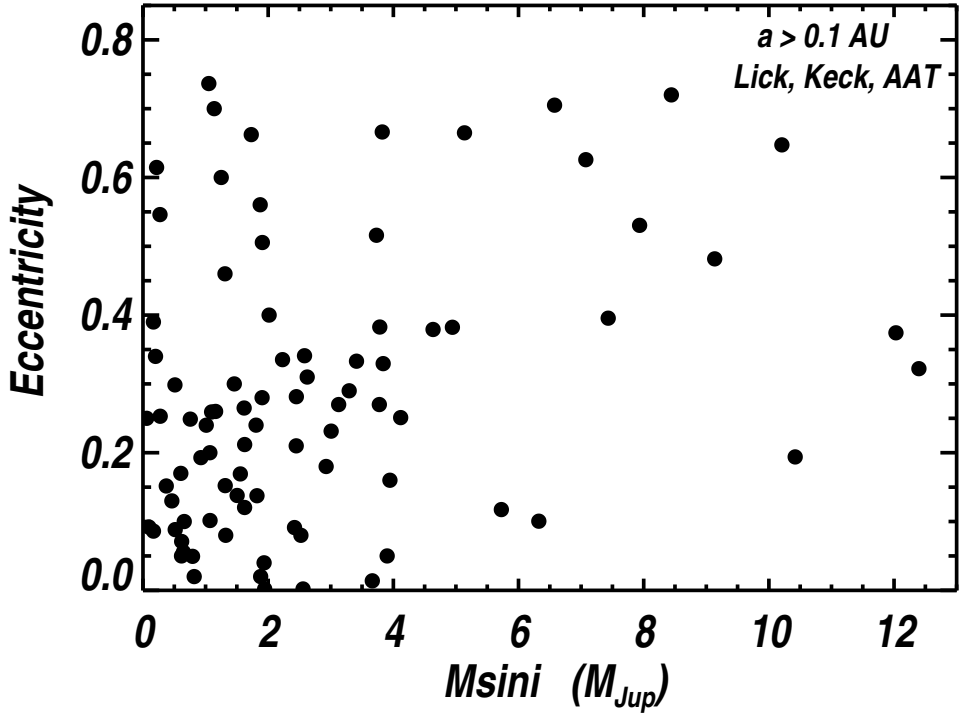}
\caption{Orbital eccentricity vs Planet mass ($M{\sin i}$) for the 104 planets
found in our survey.  A strong correlation is not seen.   However,
planet of highest mass tend to have higher orbital
eccentricities than those of lower mass.  This is puzzling,
as higher mass planets require greater perturbations to alter
orbits that were originally circular.}
\end{figure}

\subsection{Planet-metallicity correlation}
Planet occurrence correlates strongly with the abundance of heavy
elements in the host star, as shown in Fig.~6. In our survey of FGK
stars, $\sim$25\% of the most metal-rich stars, [Fe/H] $>$ +0.3,
harbor planets while fewer than 3\% of the metal poor stars, [Fe/H]
$<$ $-0.5$ have detected planets \cite{Fischer05, Santos04a, Reid02,
Gonzalez97}.  

A power-law fit to the occurrence of planets as a function
of [Fe/H] yields

\begin{displaymath}
{\cal P}{\rm(planet) = {0.03 \times \Biggl({(N_{Fe}/N_H) \over (N_{Fe}/N_H)_\odot}\Biggr)^{2}}}.
\end{displaymath}

Apparently the occurrence of gas giant planets is nearly proportional
to the square of the number of iron atoms.  This is consistent with
collision rates, suggesting that the dust particle growth rate in the
protoplanetary disk is related to the final existence of a gas giant
planet.  This steep dependence of planet occurrence on metallicity
lends weight to the core accretion model.

The physical mechanism for the observed planet-metallicity correlation
is often cast as ``nature or nurture''.  In the former case, high
metallicity enhances planet formation because of increased
availability of small particle condensates, the building blocks of
planetesimals \cite{Kornet05}.  In the latter case, enhanced stellar
metallicity is due to the late-stage accretion of gas-depleted,
dust-rich material, causing ``pollution'' of the star's convective
zone (CZ).  These two mechanisms leave different and distinugishable
marks on the host stars.  In the former case, the star is metal-rich
throughout its interior.  In the latter case, additional metals are
mixed from the photosphere only throughout the convective zone,
leaving the interior of the star with lower metallicity.

There is strong support for the former ``nature'' hypothesis. Of
particular importance, the metallicities of stars are independent of
both CZ depth and of evolution across the subgiant branch (where
dilution is expected due to a deepening CZ).  While accretion of
metals must occur for all pre-main-sequence stars, the stars with
extrasolar planets appear to have enhanced metals extending below the
CZ.  Thus it is unlikely that the high metallicity of planet-bearing
stars is caused by accretion.  Furthermore, planet-bearing stars with
super-solar metallicity are more than twice as likely to have multiple
planet systems than planet-bearing stars with sub-solar metallicity.
Taken together, these findings suggest that initial high metallicity
enhances planet formation, providing support for the core accretion
model of giant planet formation.

\begin{figure}[t]
\includegraphics[width=14cm,clip]{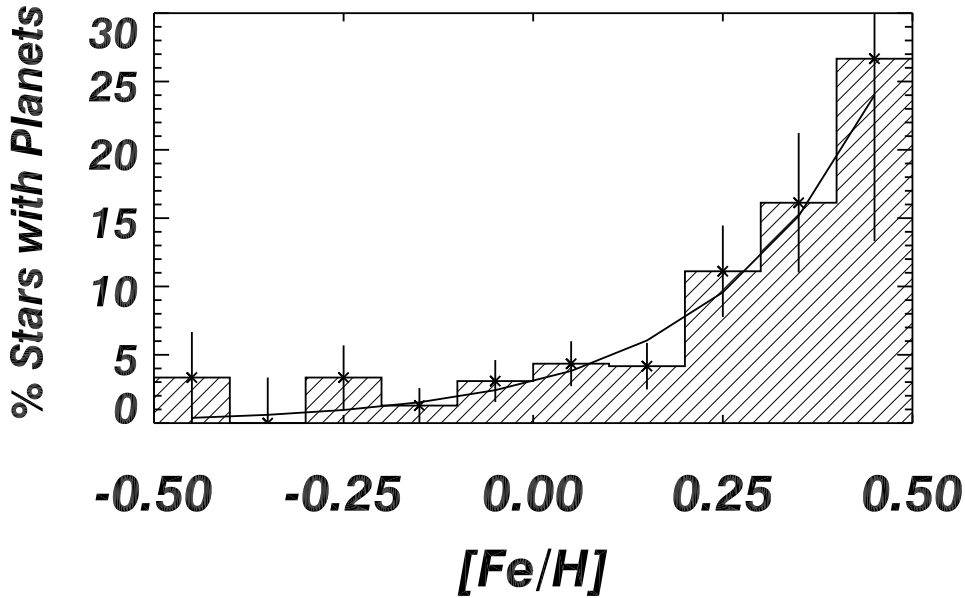}
\caption{The occurrence of exoplanets vs iron abundance [Fe/H] 
of the host star measured spectroscopically \cite{Fischer05}.
The occurrence of observed giant planets increases strongly
with stellar metallicity. The solid line is a power law
fit for the probability that a star has a detected
planet: ${\cal P}{\rm(planet) = 0.03 \times 10^{2.0 \times [Fe/H]}}$ }
\end{figure}

\section{Multi-planet systems and the lowest mass planets}

Among 152 exoplanets found by all Doppler teams, 
131 stars harbor a single planet, 14 stars have
two known planets, two have three known planets (Upsilon And and
HD~37124), and one has four detected planets (55 Cancri).  Thus,
multi-planet systems exist in 17 of 131 (13\%) of known planet-bearing
stars.  This fraction is certainly a lower limit, as multiple systems
demand more Doppler observations to extract the multiple signals.
The 17 known multi-planet systems are shown in Fig.~7, with
each planet shown at its semimajor axis and with a dot diameter
proportional to $M{\sin i}$.   

There is preference for the lower mass planet to reside inward of the
outer planets.  Such an effect is expected if the disk accretion onto
the inner planets is blocked by accretion onto the outer planet
\cite{Kley05}.  However this modest mass difference between inner and
outer planets could also be a selection effect. Outer planets that are
lower mass can be absorbed in the model of a single, more massive
planet located closer to the host star.  Additional planets continue
to emerge in the set of stars with known planets as more observations
are obtained and as Doppler precision improves.

\begin{figure}[t]
\includegraphics[width=14cm]{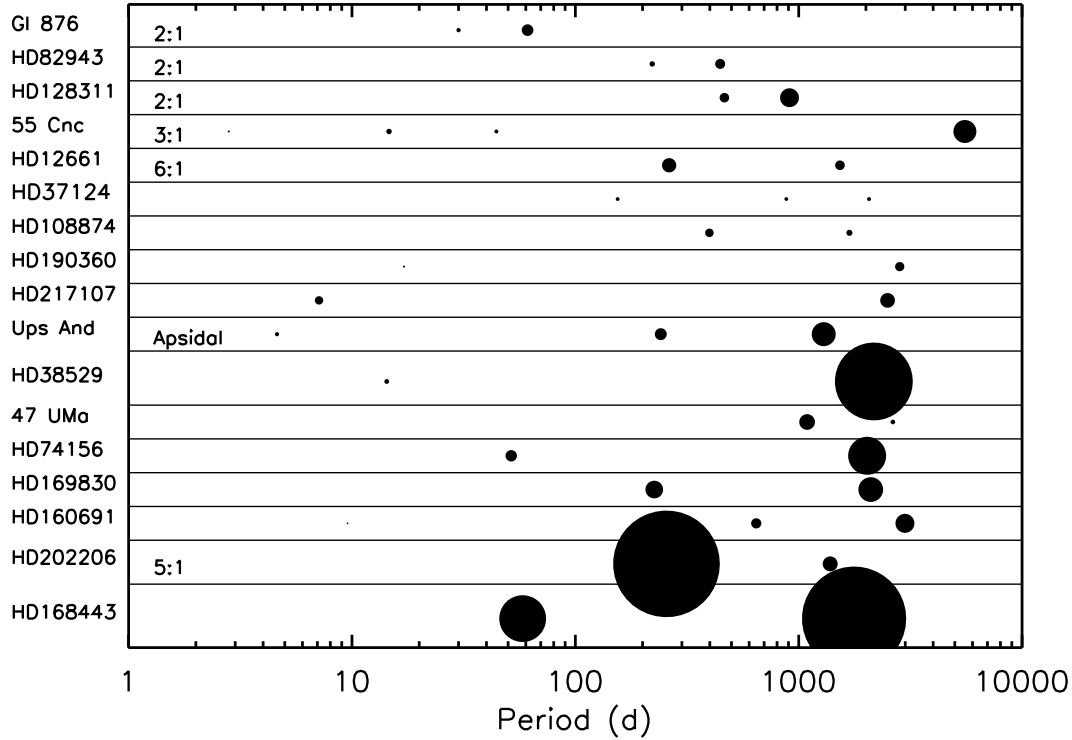}
\caption{The known 17 multi-planet systems.   The dots mark the
  orbital period of each planet and the diameter of each dot is
  proportional
to $M{\sin i}$.  There is a modest tendency for the inner planets of
  multi-planet systems to be the least massive, interpretable either
as suppressed accretion from the outer disk, or as a mere selection effect.}
\end{figure}

We have found three ``Neptune-class'' planets with minimum masses of
21, 15, and 18, $M_{\rm Earth}$, all with short periods of 2.5--10 d
around host stars, GJ 436, 55 Cancri, and HD 190360, respectively
\cite{Butler04,McArthur04,Vogt05}.  We have also found the first
``super-Earth'', with M$\sin i$ = 6.0 M$_{\rm Earth}$.  Its orbital
period is $P$=1.94 d around the star GJ 876, joining its two resonant
Jupiter-mass planets.  Apparently, planets exist that are intermediate
in mass between the ice giants and the terrestrial planets.  Such
planets may form in dusty disks that have little gas.

The Doppler method with state-of-the-art precision of 1 \mbox{m
s$^{-1}~$} can reveal planets having masses as low as 10 $M_{\rm
Earth}~$ for periods less than 5 d.  But astrophysical noise
(``jitter'') caused by stellar surface turbulence, spots, and stellar
acoustic $p$-modes make the detection of planets below 10 $M_{\rm
Earth}~$ difficult, notably due to the unpredictable, stochastic
interference of the acoustic $p$-modes in Solar-type stars.

The Doppler detection of Earth-mass planets orbiting a Solar-mass star
at $\sim$1 AU will require a 6-meter class, dedicated telescope to
detect the Doppler amplitude of only $\sim$0.1 \mbox{m s$^{-1}~$}.  An
alternative approach will be to search for low-mass planets by
achieving 1 \mbox{m s$^{-1}~$} precision for much lower-mass stars.
Surveys of M- and L-dwarfs with masses ~0.1 $M_{\odot}$ at 1 \mbox{m
s$^{-1}~$} would also allow the the detection of sub-10 $M_{\rm
earth}$ planets, though such searches will require the use of either
much larger telescopes in the optical (where these stars are very
faint), or a new generation of near-infrared echelle spectrographs on
8m-class telescopes.

The Kepler and COROT missions are designed to photometrically detect
Earth-mass planets during transits of the host star, providing the
first measure of the occurrence of rocky planets and ice-giants.
However, the host stars will reside at typical distances beyond 250
pc, making imaging and spectroscopic follow-up of the planets
difficult. A method is needed to detect earth-mass planets around
nearby stars, amenable to follow-up.

\section{The Space Interferometry Mission}

The Space Interferometry Mission, SIM, will do astrometry by using a
9-meter baseline and optical wavelengths to measure the optical path
delay. The precision for stars brighter than $V=10$ will be 1.5 $\mu$as.
SIM will carry out Galactic and extragalactic projects that require
high astrometric precision, 1.5--20 $\mu$as, during its nominal
five-year mission starting in 2011.  SIM will carry out a search for
rocky planets around $\sim$250 stars located within 20 pc.

The technical specifications of SIM are provided by \citen{Shao03}.
SIM will be carried into an Earth-trailing solar orbit via an
expendable launch vehicle and will slowly drift away from the Earth at
0.1 AU/yr, reaching 0.6 AU after 5.5 years.  It will obtain fringes at
a set of wavelengths from 400--900 nm, from which optical path delays
will be measured.  The 9 meter baseline vector between the two mirrors
is established by a separate guide interferometer that monitors bright
stars.

The astrometry of each target star is carried out relative to at least
3 reference stars located within $\sim$1 deg.  Ideally, the three
reference stars should spatially encompass the target star to
constrain the differential angular separations in both of two axes.
Ideally, the reference stars are bright ($V < 10$) and distant ($\sim$1
kpc) giants so that the astrometric ``noise'' due to planets around
them is minimized.  Candidate reference stars having brown dwarf or
stellar companions within 10 AU are identified and rejected by using
repeated (and ongoing) radial velocity measurements that we acquire
from ground-based telescopes at a precision of 20 \mbox{m s$^{-1}~$}
\cite{Frink01}.  At $V=10$ mag, a ten-chop sequence between target and
each reference star, with 30 sec integrations per chop, will achieve
the 1.5 $\mu$as \ precision, including instrumental and
photon-limited errors.  Spots on the youngest, most active, stars will
move their photocenters by $\sim 1 \mu$as, but stars older than 2 Gyr
have spot covering factors less than 0.2\%, making the spots
insignificant.

\subsection{Finding earth-mass planets with SIM}

The SIM search for rocky planets around nearby stars includes
$\sim$250 AFGKM-type targets within 20 pc.  Several criteria governed
the selection of target stars, including proximity and large angular
separation of their habitable zones.  The highest priority members of
the target list are listed at
\begin{verbatim} 
http://www.physics.sfsu.edu/SIM/
\end{verbatim}

SIM will detect planets with masses greater than 3 $M_{\rm Earth}~$
orbiting between 0.1 and 2 AU around nearby stars. It will determine
the masses, orbits, and multiplicity of planets, and will permit
correlation of these properties with the star's mass and metallicity.
Rocky planets detected by SIM will be separted from the host star by
$\sim$0.3 arcsec, offering opportunities for surgical follow-up
observations by existing ground-based and space-borne telescopes.
Thus, SIM is expected to initiate an era of characterization of rocky
planets.  Moreover, the SIM results provide reconnaissance for later
imaging missions such as TPF and Darwin.

\subsection{The lowest mass planets detectable by SIM}

The detection thresholds by SIM have been assessed by
 24), 73) and 74) with special attention paid
to secure detections of rocky planets.  Here, we consider the
detectability by SIM of planets having the lowest detectable masses of
3 $M_{\rm Earth}$ and below.

We simulate SIM measurements of a planet in a circular orbit at 1 AU
orbiting a star of 0.7 $M_{\odot}$ at 5 pc, typical of stars on the
SIM target list and the nearest stars on the TPF target lists.  We
adopt an assumed inclination of $i=60$ deg which acts to suppress the
astrometric signal in one dimension.  We consider two cases of 30 and
50 SIM observations (in each of two dimensions) obtained during 5
years with expected astrometric errors of 1.5 $\mu$as.

The timing of the simulated observations was semi-random with a
minimum separation in observations of 30 days because the lowest mass
planets cannot be detected in short periods.  In any case, detection
efficiency is not a sensitive function of the cadence of observations
\cite{Fordpasp04}.  We imposed gaps in the observing sequence to account
for the 4 month sun-avoidance time which introduces detection blind
spots near periods of 1.0 and 0.5 years.  The current key
planet-search projects have been allocated sufficient time for 24 2-d
observations of $\sim$135 stars.  However, allocation of addional
observing time is being requested and adaptive scheduling algorithms
may allow observations to be timed to optimize detections after early
data are acquired \cite{Ford05}.


Detection of 3 $M_{\rm Earth}~$ planets around such stars
is challenging because the astrometric wobble of 2.5 $\mu$as
is only slightly larger than SIM errors, 1.5 $\mu$as.  However, the
points carry a temporal coherence with orbital phase, making a
detection possible.  One detection approach is to fit the data with a
Keplerian model to obtain $\chi^2$ and determine the associated False
Alarm Probability (FAP) by using Keplerian fits to mock velocity sets that
contain no planet.

\begin{figure}
\includegraphics[width=14cm]{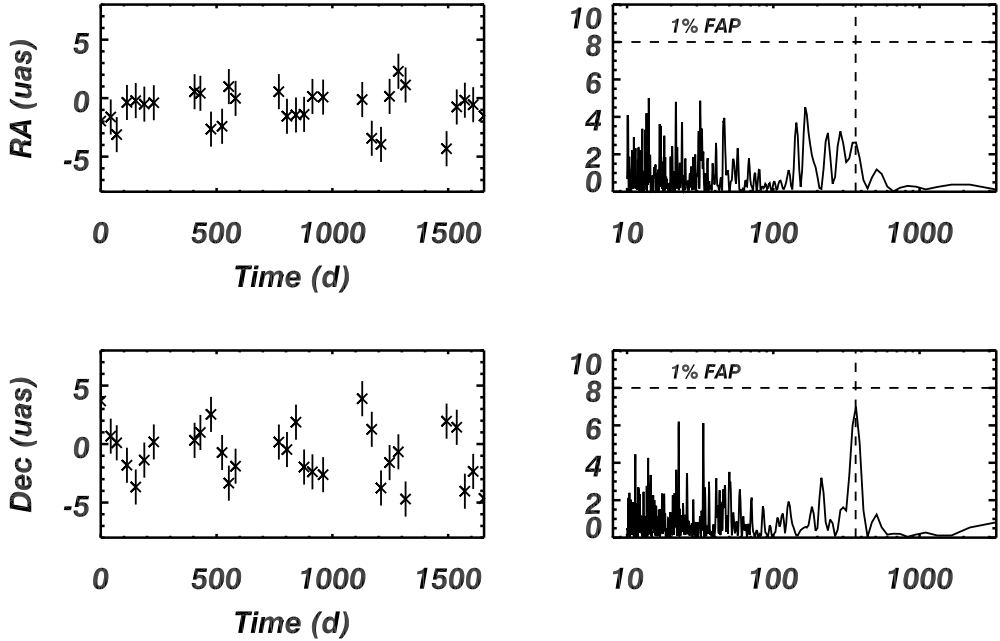}
\caption{Simulation of 30 SIM measurements in orthogonal directions
  (labeled RA and DEC) for a 3 $M_{\rm Earth}$ planet orbiting at 1 AU
  from a 0.7 $M_{\odot}$ star at 5 pc.  The periodicity in the DEC
  measurements (long axis) is marginally apparent.  The periodogram of position
  for the RA and DEC measurements is shown at right.  For the DEC
  measurements, the signal has a FAP of just above 1\%, implying that
  planets of 3 $M_{\rm Earth}$ at 1 AU are marginally detectable (star at 5 pc).}
\end{figure}
\begin{figure}[!ht]
\includegraphics[width=14cm]{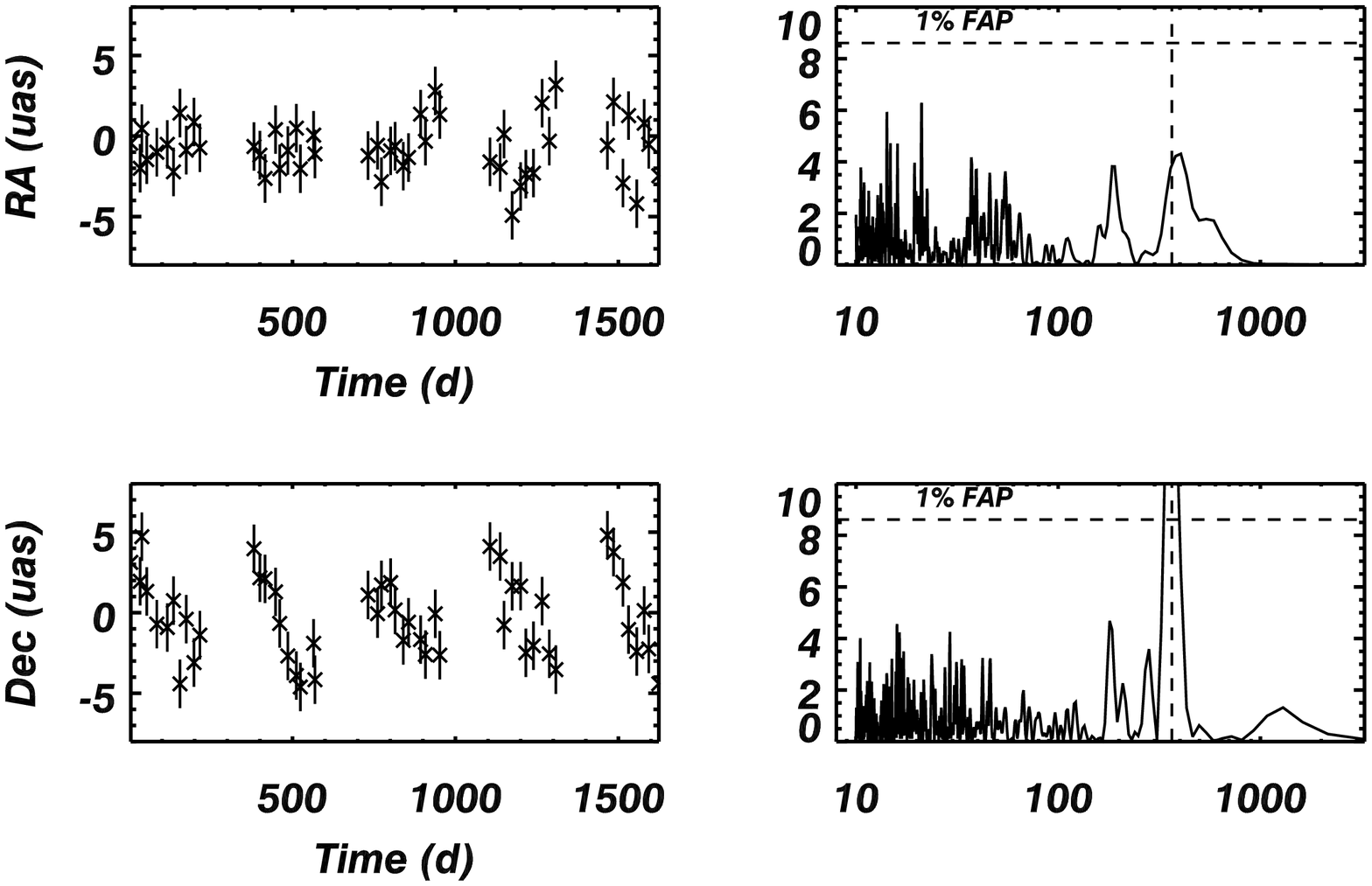}
\caption{Same as Fig.~8, but for 50 observations.
The 3 $M_{\rm Earth}$ stands out strongly in the periodogram,
due to the extra observations.
}
\end{figure}

Here, we simplify the estimate of FAP by computing a periodogram of
the astrometric measurements along both axes (labelled ``RA'' and
``DEC'') and determining the FAP by Monte Carlo of data sets that have
no planet.  Figure 8 shows (at left) a typical set of simulated
astrometric measurements (due to the 3 $M_{\rm Earth}~$ planet and 1.5
$\mu$as \ noise) in both RA and DEC, the latter showing a marginal
periodicity to the eye (the arbitrary inclination supresses the signal
in RA).  At right in Fig.~8, the periodogram of the astrometric
measurements shows a peak residing just below the 1\% FAP threshold.
{\em Thus, with only 30 observations SIM can just detect planets
of 3 $M_{\rm Earth}~$ orbiting at 1 AU around Solar type stars at 5
pc.}  

If 50 SIM observations are made, the astrometric periodicity in
the signal stands out strongly, as shown in Fig.~9. Thus, there is a
steep improvement in the detectability of planets of 3 $M_{\rm
Earth}~$ by increasing the number of observations from 30 to 50, due
to signal being comparable to the errors. These results are similar to
those of Sozzetti et al. \cite{Sozzetti02} who assumed somewhat
fewer observations per star.

We have run 1000 realizations of the two cases, 30 and 50 observations
for this same case of a 3 $M_{\rm Earth}~$ planet orbiting at 1 AU
around 0.7 $M_{\odot}$ star at 5 pc.  The cumulative distribution
function of FAP values from the trials is shown in Fig.~10.  The
distribution shows that with 50 observations, planets of 3 $M_{\rm
Earth}~$ are easily detected and carry low FAP, typically below 0.01.
However, if only 30 observations are made, the FAP is typically 0.04
or above in order for 80\% of these planets are to be detected.  Even
with only 30 observations, SIM is capable of detecting planets of 3
$M_{\rm Earth}~$ with 63\% efficiency, incurring a FAP of only
$\sim$2\%.  SIM can also identify stars highly likely to harbor
planets of slightly lower mass, but with successively higher FAP.

\begin{figure}[!ht]
\includegraphics[width=14cm]{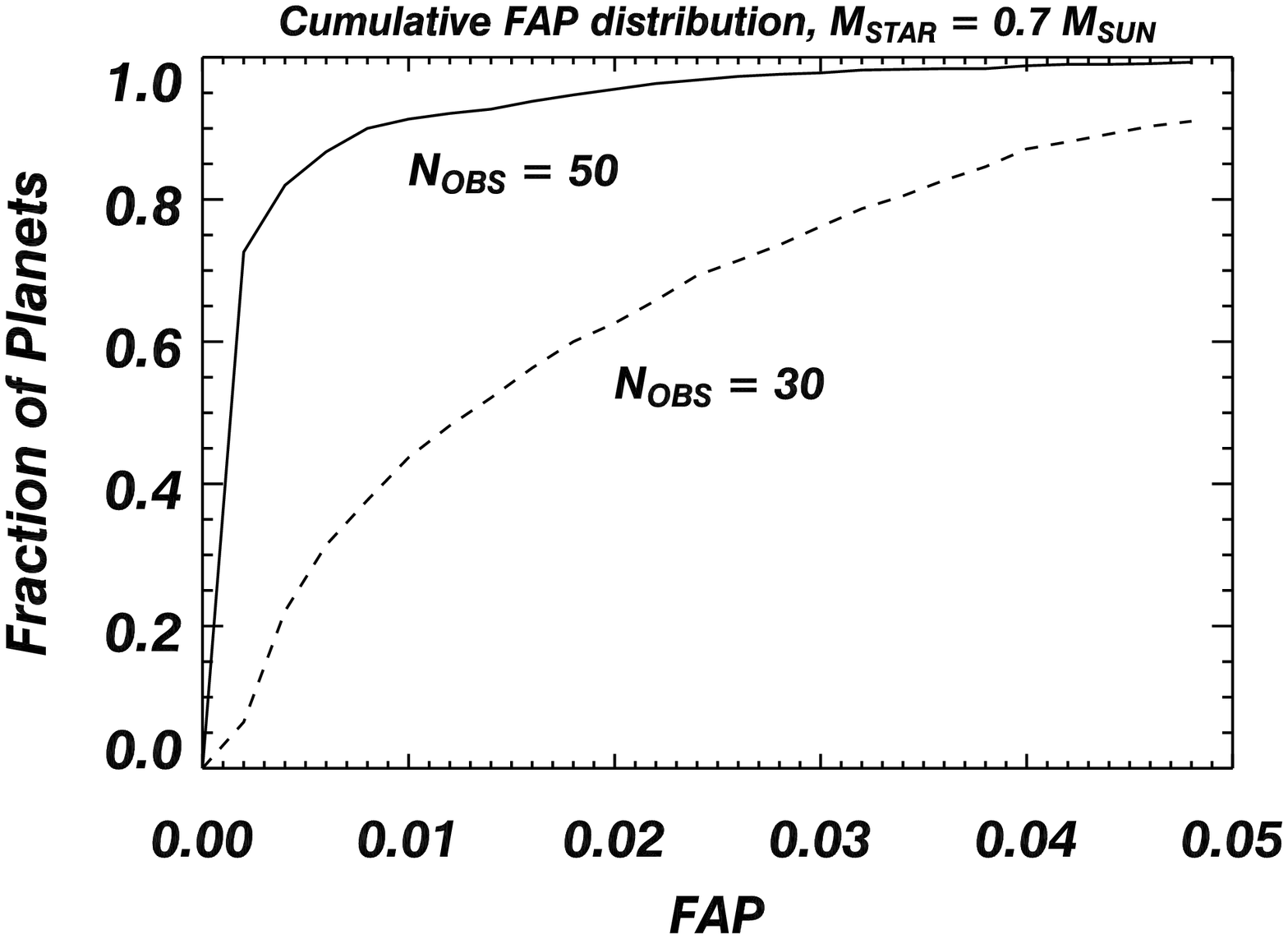}
\caption{The fraction of 3 $M_{\rm Earth}$ planets in the Habitable
Zone that are detectable 
by SIM, as a function of the associated False Alarm Probability (FAP)
for two cases: 30 and 50 SIM observations.
The simulations adopt the nominal SIM astrometric
precision of 1.5 $\mu$as \ in each orthogonal direction,
and adopt random times of observation, excluding sun-avoidance
times.   The host star is assumed to have mass of 0.7 $M_{\odot}$,
at a distance of 5 pc, representative of nearby late G and K dwarfs.
For the case of 50 observations, 98\% of planets having 3 $M_{\rm Earth}$
are detected, for an adopted FAP in the signal of 0.03 (3\% of
stars will incur a false detection of a planet).   For 30 observations,
75\% of the  3 $M_{\rm Earth}$ planets will be detected at FAP = 0.03.
Thus, SIM will detect the majority of 3 $M_{\rm Earth}$ planets and incur
modest false alarms.}
\end{figure}

We also considered a 1.5 $M_{\rm Earth}~$ planet orbiting (as before)
with $P\approx 1$ yr around a 0.7 $M_{\odot}$ star at 5 pc.  For this
lower mass planet, the astrometric wobble is slightly less than the
measurement errors of 1.5 $\mu$as, making detection even more
challenging.  However, Monte Carlo trials of such a system typically
yield a weak but apparent periodogram peak corresponding to FAP =
3--5\% with only 30 measurements, as shown in Fig.~11.  {\em Thus, SIM
can marginally detect planets of 1.5 $M_{\rm Earth}~$ orbiting at $\sim$1 AU
around Solar type stars at 5 pc, albeit with considerable false alarms.}
Nonetheless these marginal detections will provide a subsample of stars
that are enriched in planets of 1.5--3.0 $M_{\rm Earth}$, useful for
follow-up work.
c
\begin{figure}
\includegraphics[width=14cm]{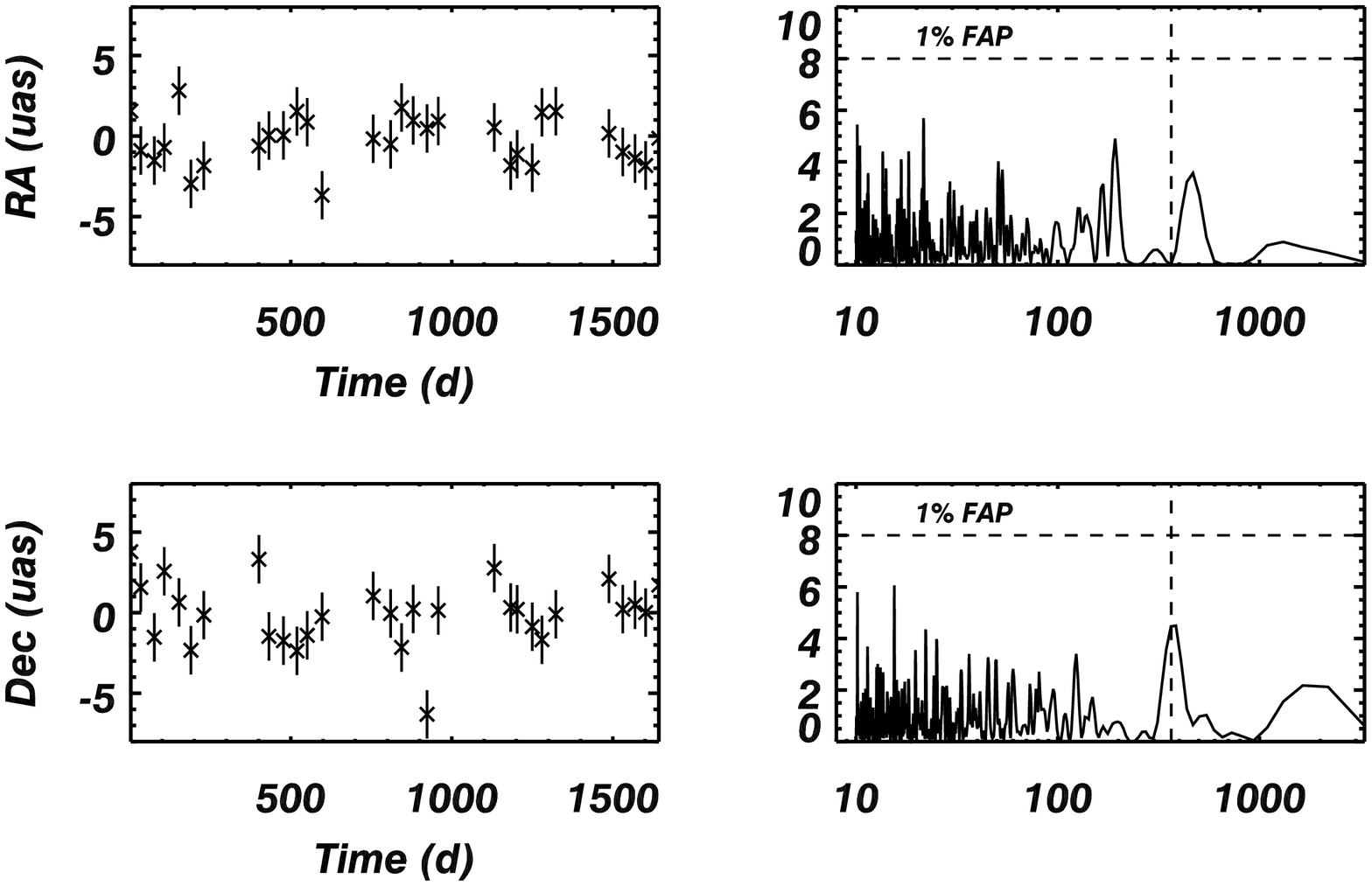}
\caption{Simulated 30 SIM measurements in orthogonal directions for a
  1.5 $M_{\rm Earth}$ planet orbiting with $P\approx$1 yr from a 0.7
  $M_{\odot}$ star.  The periodicity in the DEC measurements (long
  axis) is apparent weakly in the periodogram for the DEC measurements
  at the correct period of $P$ = 365 d, but with a FAP of 4\%, typical
  of the 100 trials.  Thus, planets of 1.5 $M_{\rm Earth}$ at 1 AU
are only  marginally detectable.}
\end{figure}

\section{The synergy of SIM and TPF/Darwin}

The simulations of SIM observations of Earth-mass planets show that 3
$M_{\rm Earth}~$ planets are detectable and 1.5 $M_{\rm Earth}~$
planets are marginally found at 5 pc.  Thus, the SIM survey of 200
nearby stars will identify a subset that has planets of 3--10 $M_{\rm
Earth}~$ (should they be common) and another subset that is likely to
have even lower mass planets, 1.5--3.0 $M_{\rm Earth}$, albeit with
some false alarm interlopers.  SIM can thus produce an input sample of
nearby stars that is enriched by about a factor of 3 in 1.5--3 $M_{\rm
Earth}~$ planets over an original sample.  Assuming, for example, that
the fraction of stars with earths in the habitable zone, $\eta_{\rm
Earth}$, is 0.1, it is easy to show that SIM will produce an output
list of stars that is enriched by a factor of 3 in habitable earths
over the original input sample of stars.  Thus, SIM will provide TPF
and Darwin with target stars having either strong or plausible
evidence of rocky planets.  SIM will also identify those stars that
TPF and Darwin should avoid, notably those with a large planet near
the habitable zone that renders any earths dynamically unstable.  Of
course, any such saturn or neptune-mass planets within 2 AU will be
valuable themselves for planetary astrophysics.

If $\eta_{\rm Earth}$ is indeed $\sim$10\%, TPF/Darwin will be hard
pressed to detect these few earths because of their rarity and
their faintness, $V\approx 30$ mag.  Moreover, for modestly inclined
orbital planes, TPF/Darwin will miss planets located angularly within
the diffraction-limited ``inner working angle'' (IWA = $\sim 4
\lambda/D$ = 0.065 arcsec for TPF-C).  A planet orbiting 1 AU from a
star located 5 pc away will spend roughly 1/3 of its orbit inside the
IWA, leaving it undetected.  {\em Thus if the occurrence of earths in
habitable zones is of order 10\%, SIM will triple the efficiency of
TPF and Darwin both by identifying the likely host stars and by
predicting the orbital phase during which the earth is farthest from
the glare of the host star.}

SIM alone provides a wealth of planetary astrophysics, including the
masses, orbital radii, and orbital eccentricities of rocky planets
around the nearest stars.  It will also find correlations between
rocky planets and stellar properties such as metallicity and rotation.
With a lifetime extended beyond 5 years, SIM can detect
planets of even lower mass, down to 1 $M_{\rm Earth}$.

SIM and TPF/Darwin together, along with Kepler, provide a valuable
combination of information about rocky planets.  Kepler offers the
occurence rate of small planets.  SIM provides the masses and orbits
of planets around nearby stars, identifying the candidate earths.
TPF/Darwin measure radii, chemical composition, and atmospheres. In
some cases, images from TPF/Darwin may feedback on the analysis of old
SIM data, helping orbit determination especially for multiple
planetary systems.

\section*{Acknowledgements} 
We thank John Johnson, Chris McCarthy, Brad Carter, and Alan Penny for
their work on the Doppler planet search.  We thank the SIM Project
team at JPL, especially Michael Shao, Chas Beichman, Steven Unwin,
Shri Kulkarni, Chris Gelino, Joe Catanzerite and Jo Pitesky.  We
appreciate support by NASA grant NAG5-75005 and NSF grant AST-0307493.


  

\end{document}